\documentclass{PoS}

\usepackage{amsmath}
\usepackage{amssymb}
\usepackage{subcaption}

\title{Fourier acceleration, the HMC algorithm and renormalizability}

\ShortTitle{Fourier acceleration, the HMC algorithm and renormalizability}

\author{\speaker{Norman H.~Christ}\thanks{This work partially supported by US DOE grant \#DE-SC0011941 and the DOE Exascale Computing Project (ECP), Project
Number: 17-SC-20-SC.}\\
       Department of Physics, Columbia University, New York, NY 10027, USA\\
        E-mail: \email{nhc@phys.columbia.edu}}
\author{Evan W.~Wickenden\\
         Department of Physics, Columbia University, New York, NY 10027, USA\\
        E-mail: \email{eww2119@columbia.edu}}

\abstract{The analysis developed by L\"uscher and Schaefer of the Hybrid Monte Carlo (HMC) algorithm is extended to include Fourier acceleration.  We show for the $\phi^4$ theory that Fourier acceleration substantially changes the structure of the theory for both the Langevin and HMC algorithms.  When expanded in perturbation theory, each five-dimensional auto-correlation function of the fields $\phi(x_i, t_i)$, $1\le i \le N $, corresponding to a specific 4-dimensional Feynman graph separates into two factors: one depending on the Monte-Carlo evolution times $t_i$ and the second depending on the space-time positions $x_i$.  This separation implies that only auto-correlation times at the lattice scale appear, eliminating critical slowing down in perturbation theory.}

\FullConference{The 36th Annual International Symposium on Lattice Field Theory - LATTICE2018\\
		22-28 July, 2018\\
		Michigan State University, East Lansing, Michigan, USA.}

\begin{document}

\section{Introduction}

One of the most challenging aspects of numerical studies using lattice QCD is generating  stochastic samples of gauge configurations distributed according to the weight required by the QCD path integral.  The current best method for constructing such a sample is the Hybrid Monte Carlo (HMC) algorithm~\cite{Duane:1987de}.  In this method these samples are obtained as elements of a Markov chain generated by a sequence of molecular dynamics evolution steps followed by a Langevin-style, Gaussian refresh of the evolution momenta.  The molecular dynamics follows conventional Hamiltonian mechanics with the gauge matrices treated as classical coordinates and fictitious momenta conjugate to gauge variables defined in the Lie algebra of the gauge group .  

These conjugate momenta enter the evolution Hamiltonian through a simple ``kinetic'' energy term $\frac{1}{2} \sum_{\mu,x,b} (P^b_\mu(x))^2$ where $b$, $1 \le b \le 8$, labels a generator of the $SU(3)$ gauge group and the position $x$ and direction $\mu$ identify a lattice link.  The scale of molecular dynamics time is chosen so that the ``mass'' appearing in this Hamiltonian kinetic energy is unity.   As calculations are performed at increasingly small lattice spacing the frequencies present in this classical system extend over a larger range corresponding to an increasingly broad distribution of forces. Since this kinetic energy term implies a single distribution of velocities, the molecular dynamics evolution encounters critical slowing down.  The integration of the equations of motion must be carried out with a small step size to allow an accurate evolution for those modes with very large forces, typically ultraviolet, lattice-scale degrees of freedom.  This implies that the more physically important long-distance modes which move at the same velocity but for which the potential changes more slowly, must be evolved for an increasing number of time steps to generate an independent sample.

In simpler systems, this critical slowing down of the classical evolution can be mitigated by introducing a mode-dependent mass chosen to give a larger velocity for those modes with smaller frequency, a technique referred to as Fourier acceleration.  Such an approach is challenging in a gauge theory where local gauge symmetry implies that simple Fourier transformation cannot be used to distinguish the soft and stiff modes.  However, there are a number of interesting proposals that allow Fourier acceleration to be applied to gauge theories~\cite{Girolami:2011xx,Cossu:2017eys,Jin:2018xxx,Zhao:2018xxx}.

Apart from the difficulty of local gauge symmetry, the asymptotic freedom of QCD might suggest that this is a highly promising application for Fourier acceleration since at short distances in a fixed gauge the modes of the theory become independent harmonic oscillators with known frequencies allowing Fourier acceleration to be applied precisely as the continuum limit is approached.  However, this is thrown into doubt because of the complexities of quantum field theory.  While the correlation functions in both space-time and Monte Carlo time for the less efficient Langevin algorithm can be analyzed in perturbation theory and shown to possess a well-defined continuum limit~\cite{ZinnJustin:1986eq} the more efficient HMC algorithm contains non-renormalizable singularities implying the presence of non-universal long-time correlations between Monte Carlo samples~\cite{Luscher:2011qa}.

As we will show, adding Fourier acceleration to the Langevin or HMC algorithms substantially changes both algorithms and removes correlations between Monte Carlo samples separated by physical times.  Thus, Fourier acceleration renders both algorithms ``renormalizable'' in the trivial sense that there are no Monte Carlo time correlations at a  physical-scale in the continuum limit.  Since the Langevin algorithm is simpler to analyze than the HMC we will discuss them both with an emphasis on Langevin evolution.

\section{Stochastic evolution}

Following the treatment of Luscher and Schaefer (L\&S) we will examine a $\phi^4$ theory.  For this case the Langevin equation determines the evolution of the real classical field $\phi(x,t)$ as a function of Monte Carlo time $t$ and as a functional of the noise field $\eta(x,t)$:
\begin{equation}
\partial_t\phi(x,t) = -\frac{\delta}{\delta(x,t)}S[\phi(x,t)] + \eta(x,t)
                           = -(-\partial^2 + m_0^2)\phi(x,t) - \frac{g_0}{3!}\phi(x,t)^3 +\eta(x,t)
\label{eq:L}
\end{equation}
where $x$ is a space-time coordinate.  At large evolution time $t$, the resulting stochastic field $\phi(x,t)$ will be distributed according to the desired path integral weight $e^{-S[\phi(t)]}$ if the noise field obeys :
\begin{equation}
\left\langle\eta(x,t) \eta(y,s)\right\rangle = 2 \delta^4(x-y)\delta(t-s).
\label{eq:L-noise}
\end{equation}
The same distribution results from an ensemble $\phi(x,t_n)$ obtained from a single noise field collected at different sample times $t_n$.  One may view this as a useful algorithm for generating stochastic path integral samples~\cite{Batrouni:1985jn} or as a new formulation of quantum mechanics based on Brownian motion~\cite{Parisi:1980ys}.

\begin{figure}[h]
\centering
\includegraphics[width=0.8\textwidth,angle=0]{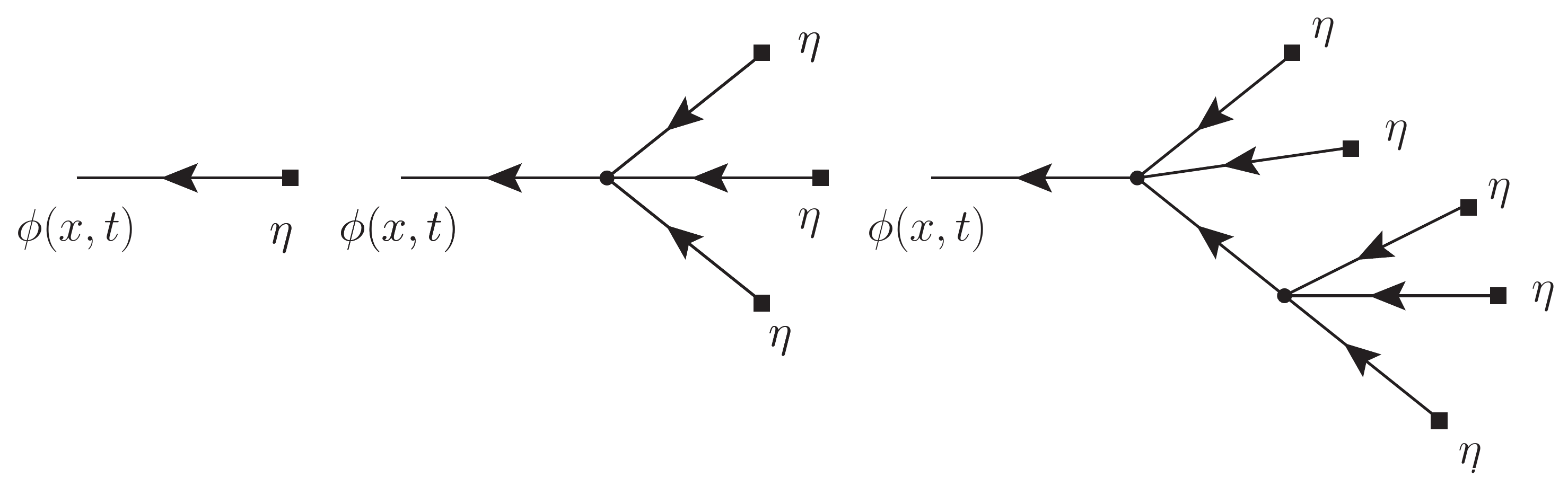}
\caption{Graphical representation of terms that would appear in a recursive solution of Eq.~\eqref{eq:L-IE}.}
\label{fig:expansion}
\end{figure}

\begin{figure}[h]
\centering
\includegraphics[width=0.575\textwidth,angle=0]{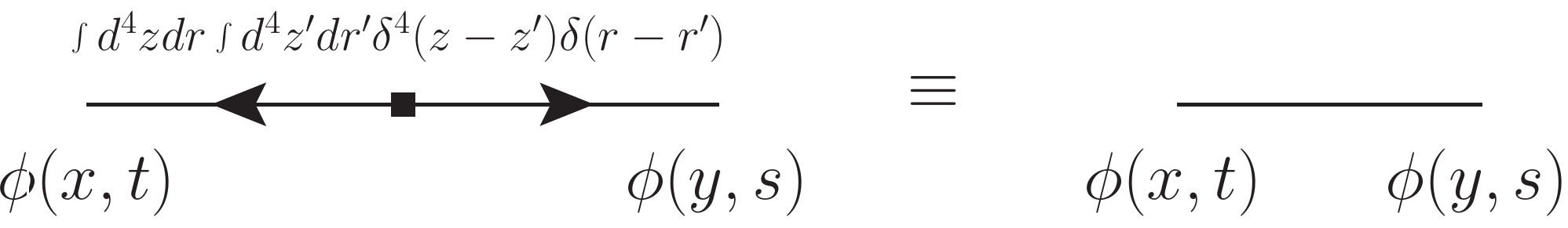}
\caption{Graphical version of the process of averaging the product of two noise sources and representing the resulting product of directed propagators by a single directionless line.}
\label{fig:line}
\end{figure}

At large time Eq~\eqref{eq:L} can solved as a power series in $g_0$ by iterating the equation:
\begin{equation}
\phi(x,t) = \int d^4y \int d s K_{\mathrm{L}}(x-y, t-x) \left\{\eta(y,s) - \frac{g_0}{3!}\phi(x,t)^3\right\}
\label{eq:L-IE}
\end{equation}
using the kernel function $K_{\mathrm{L}}(x,t)$ given by
\begin{equation}
K_{\mathrm{L}}(x,t) =  \theta(t)\int d^4 q \frac{e^{iq \cdot x}}{(2\pi)^4}e^{-(q^2+m_0^2)t}.
\label{eq:L-K}
\end{equation}
Iteratively replacing the fields $\phi$ in the $g_0\phi^3$ term on the right hand of Eq.~\eqref{eq:L-IE} by solutions obtained in lower order will generate a series expansion, giving the field $\phi$ as a function of $\eta$ which can be described by a graphical expansion of the sort shown in Fig.~\ref{fig:expansion}.  The perturbative result for a general Green's function would be obtained by multiplying an appropriate number of such solutions $\phi(x_1,t_1)\phi(x_2,t_2)\ldots\phi(x_N,t_N)$ and then averaging over the noise.  The noise functions will combine in pairs.  This will join pairs of directed lines ending in a square into what, following the conventions of Ref.~\cite{Luscher:2011qa}, we would represent as a single, directionless line, as suggested in Fig.~\ref{fig:line}.

In contrast to the Langevin algorithm, the HMC involves infrequent refreshes of the molecular dynamics momentum and relies for its efficiency on a Hamiltonian molecular dynamics trajectory during which the system moves deterministically to a new point in which all modes have changed substantially from their starting point.  Ideally, if the initial momenta are chosen randomly, the start and end points of this trajectory will be largely independent. Thus, it is this single-trajectory evolution which we wish to Fourier accelerate.  

Following L\&S, such a classical evolution can be analyzed if an additional stochastic element is introduced by using the Generalized HMC algorithm (GHMC)~\cite{Horowitz:1991rr}.  The analogue to the Langevin equation is a pair of Hamiltonian-like equations,
\begin{eqnarray}
\partial_t \pi(x,t) = -\frac{\delta}{\delta(x,t)}S[\phi(x,t)] -2\mu_0 \pi + \eta'(x,t) \quad 
\partial_t\phi(x,t) = \pi(x,t),
\end{eqnarray}
which can be rewritten as the second-order evolution equation:
\begin{eqnarray}
\partial_t^2\phi(x,t) = -\frac{\delta}{\delta(x,t)}S[\phi(x,t)] -2\mu_0 \partial_t\phi(x,t) + \eta'(x,t)
\label{eq:HMC}
\end{eqnarray}
where the noise $\eta'(x,t)$ is a simple multiple of the earlier $\eta(x,t)$: $\eta'(x,t) = \sqrt{2\mu_0}\eta(x,t)$.

As discussed by L\&S the second order Eq.~\eqref{eq:HMC} can also be solved using a kernel function:
\begin{equation}
K_{\mathrm{HMC}}(x,t) = \theta(t) \int d^4 q \frac{e^{iq \cdot x}}{(2\pi)^4} e^{-\mu_0 t}
\frac{\sin(\epsilon_q t)}{\epsilon_q}
\label{eq:HMC-K}
\end{equation}
where $\epsilon_q = \sqrt{q^2+m_0^2-\mu_0^2}$.  Equations~\eqref{eq:L-K} and \eqref{eq:HMC-K} give insight into the autocorrelation times expected for Langevin and HMC evolutions.  As can be seen from the exponential  $t$ dependence of $K_{\mathrm{L}}(x,t)$ in Eq.~\eqref{eq:L-K} the longest autocorrelation times behave as $1/m_0^2$ suggesting a dynamical critical exponent of two.  For the HMC algorithm the kernel function $K_{\mathrm{HMC}}(x,t)$ does not fall exponentially as $t$ grows (for $\mu_0\to0$ where the GHMC becomes the HMC algorithm) but instead oscillates with a longest period given by $1/m_0$ suggesting a dynamical critical exponent of one.

\section{Fourier acceleration}

Fourier-accelerated versions of the Langevin algorithm can be obtained by introducing a factor of the Klein-Gordon operator into the left-hand side of Eq.~\eqref{eq:L}~\cite{Batrouni:1985jn}.  For the HMC algorithm the Klein-Gordon operator is introduced as a mass term in the fictitious HMC kinetic energy: $\frac{1}{2}\pi^\dagger\pi \rightarrow \frac{1}{2}\pi^\dagger\frac{1}{(-\partial^2 +m_0^2)}\pi$.  For our $\phi^4$ example, the resulting stochastic evolution equations become:
\begin{eqnarray}
(-\partial^2+m_0^2)\partial_t\phi(x,t) &=& -(-\partial^2+m_0^2)\phi -\frac{g_0}{3!}\phi^3
 + \widetilde{\eta}(x,t)
\label{eq:L-FA1} \\
(-\partial^2+m_0^2)\partial_t^2\phi(x,t) &=&  -(-\partial^2+m_0^2)\phi -\frac{g_0}{3!}\phi^3 -2\mu_0 (-\partial^2+m_0^2)\partial_t\phi(x,t)+\widetilde{\eta}'(x,t)
\label{eq:HMC-FA1}
\end{eqnarray}
where as before $\widetilde{\eta}'=\sqrt{2\mu_0}\widetilde{\eta}$.  However, $\widetilde{\eta}$ obeys:
\begin{equation}
\left\langle\widetilde{\eta}(x,t) \widetilde{\eta}(y,s)\right\rangle = 2 (-\partial^2 + m_0^2)\delta^4(x-y)\delta(t-s).
\label{eq:L-FA-noise}
\end{equation}

The 5-D structure of these equations has been dramatically changed by Fourier acceleration.  This can be easily seen if both equations are divided by the factor $(-\partial^2+m_0^2)$ and written as:
\begin{eqnarray}
(\partial_t + 1)\phi(x,t) &=& -\frac{g_0}{3!(-\partial^2+m_0^2)}\phi^3
 + \overline{\eta}(x,t)
\label{eq:L-FA2} \\
(\partial_t^2 +2\mu_0\partial_t +1)\phi(x,t) &=& -\frac{g_0}{3!(-\partial^2+m_0^2)}\phi^3 +\overline{\eta}'(x,t)
\label{eq:HMC-FA2}
\end{eqnarray}
where  $\overline{\eta}'=\sqrt{2\mu_0}\overline{\eta}$ and $\overline{\eta}$ obeys
\begin{equation}
\left\langle\overline{\eta}(x,t) \overline{\eta}(y,s)\right\rangle = \frac{2}{(-\partial^2 + m_0^2)}\delta^4(x-y)\delta(t-s).
\end{equation}
As can be seen from Eqs.~\eqref{eq:L-FA2} and \eqref{eq:HMC-FA2},  Fourier acceleration has effectively separated the dependence on the space-time and stochastic time variables.  Now the kernels that can be used to solve these equations perturbatively do not depend on the space-time physics.  They are given by:
\begin{equation}
K^{\mathrm{FA}}_{\mathrm{L}}(x,t) = \delta^4(x) \theta(t) e^{-t}
\mbox{  and   } 
K^{\mathrm{FA}}_{\mathrm{HMC}}(x,t) = \delta^4(x) e^{-\mu_0 t} \theta(t)\sin\bigl(t(1-\mu_0^2)^{1/2}\bigr)/(1-\mu_0^2)^{1/2}.
\end{equation}
All critical slowing down has been removed and both the auto-correlation time for the Langevin algorithm and the oscillation period for the HMC algorithm are now simply one and $2\pi$ in lattice units.  If we evaluate a particular term in the perturbative solution to either equation corresponding to a given diagram constructed from the ingredients found in Figures~\ref{fig:expansion} and \ref{fig:line} the resulting amplitude will factor into the space-time Green's function that would result if the diagram were viewed as a standard Feynman diagram multiplied by a function of the evolution times $t_i$ which is combination of exponential or trigonometric functions of the $t_i$ multiplied by a finite-order polynomial in the $t_i$.  

This can be expressed algebraically by the relation
\begin{equation}
\Bigl\langle \phi(x_1,t_1)[\overline{\eta}]\;\phi(x_2,t_2)[\overline{\eta}]\ldots\phi(x_N,t_N)[\overline{\eta}]\Bigr\rangle_{\overline{\eta}}
= \sum_{\{\Gamma_5\}} 
   I_{\Gamma_5}(t_1,t_2,\ldots, t_N) G_{\Gamma_4(\Gamma_5)}(x_1,x_2\ldots, x_N).  
\end{equation}
The left-hand side is the product of fields computed using the noise $\overline{\eta}$ and then averaged over $\overline{\eta}$.  The right-hand side is a sum over contributions from each 5-D diagram $\Gamma_5$, each of which can be written as a product of two factors.  The first is $I_{\Gamma_5}(t_1,\ldots,t_N)$ determined by that 5-D diagram and depending on the evolution times $t_1,\ldots, t_N$.  The second is the usual amplitude $G_{\Gamma_4(\Gamma_5)}(x_1,x_2\ldots, x_N)$ corresponding to the 4-D Feynman diagram $\Gamma_4(\Gamma_5)$ obtained from the 5-D diagram $\Gamma_5$ by treating each line as a 4-D Feynman propagator.

Thus, the $t$-dependent factors depend only on the stochastic evolution time and will be unaffected by the physical scales in the field theory problem being studied.  In this perturbative expansion the scale of the  evolution time dependence is that of the lattice cutoff.  Possible longer, physical time scales are absent.  The non-renormalizable, 5-D light-cone singularity with a logarithmically divergent coefficient found by L\&S in second order perturbation theory does not occur in this Fourier-accelerated perturbation theory.

Although we do not thoroughly understand the structure of this perturbative expansion of the 5-D correlation functions, we will comment on two features.  First, while individual diagrams that appear in the perturbative expansion of Eqs.~\eqref{eq:L-FA2} and \eqref{eq:HMC-FA2} that determines the $N$-point function
\begin{equation}
\left \langle \phi(x_1,t_1) \phi(x_2,t_2) \ldots \phi(x_N,t_N)\right\rangle
\end{equation}
can each be neatly written as the product of a conventional 4-D Feynman diagram multiplied by a factor which depends on the evolution times $t_1$, $t_2\ldots t_N$, these evolution-time dependent factors will be different for the different 5-D diagrams.  The correctness of the Fourier-accelerated Langevin or GHMC algorithm implies that when evaluated at equal evolution times, $t_i=t_j$ for all $i$ and $j$ these factors must combine to reproduce the required combinatoric factors that appear in the 4-D Dyson-Wick expansion.  (Note: given the arrows on the boson lines in the stochastic perturbation theory there will be many 5-D diagrams that will combine to give a single 4-D diagram.)  However, when the times $t_i$ are unequal this connection with the 4-D theory is lost.  In particular the divergent diagrams requiring renormalization subtractions and the corresponding diagrams which contain the appropriate counter terms will acquire different coefficients and no longer cancel when $t_i\ne t_j$.  While this behavior may raise concerns, such potentially strong autocorrelations at  short evolution times will have no effect at large evolution times given the rapidly falling or oscillating character of these coefficients.

\begin{figure}[h]
\centering
\includegraphics[width=0.75\textwidth,angle=0]{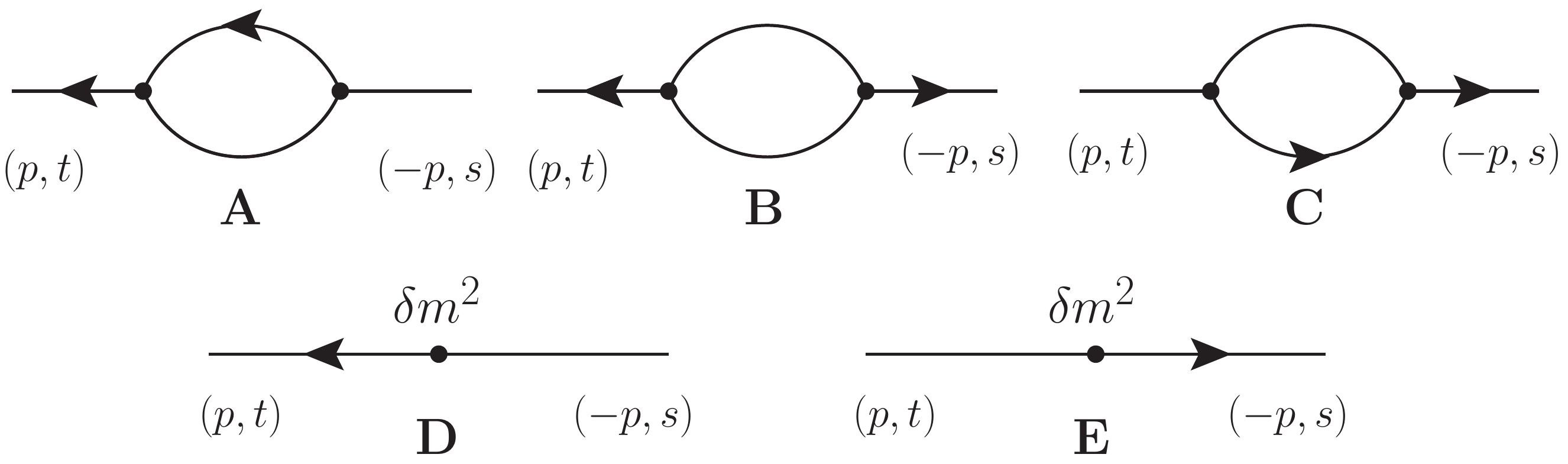}
\vskip 0.2 in
\caption{Five type of stochastic diagrams which contribute to the boson self-energy at second order in a $\phi^3$ theory.}
\label{fig:self-energy}
\end{figure}

The diagrams shown in Fig.~\ref{fig:self-energy} provide a simple example for Langevin evolution for the case of the second-order self-energy diagrams and counter terms in a $\phi^3$ theory.  The first three correspond to different factors $I_\Gamma(t-s)$ which will multiply a single conventional divergent 4-D one-loop diagram where $I_\Gamma(t-s)$ depends on the evolution times $t$ and $s$ and $\Gamma = A$, $B$ and $C$.  Similarly, the mass counter terms will be the usual 4-D counter term multiplied by amplitudes $I_D(t-s)$ and $I_E(t-s)$.  The five amplitudes $I_\Gamma$ can be easily worked out.  Setting $s=0$ we find:
\begin{eqnarray}
I_A(t)    = I_C(-t) = \frac{2}{3}e^{-|t|}\bigl[1 + \theta(t)\bigl(6t-4(1-e^{-t})\big)\bigr] \quad\quad
&&I_B(t) = \frac{2}{3}e^{-|t|}\bigl[2 - e^{-|t|}\bigr]  \nonumber \\
&& \hskip -2.7 in
I_D(t)   =e^{-|t|}\,\bigl[\frac{1}{2}+t\theta\bigl(t\bigr)\bigr] \quad\quad
I_E(t)   = e^{-|t|}\,\bigl[\frac{1}{2}-t\theta\bigl(-t\bigr)\bigr].
\end{eqnarray}
At $t=0$, $I_A(0)+I_B(0)+I_C(0) = 2\bigl(I_D(0)+I_E(0)\bigr)$ and the counter term can be chosen to  cancel the divergent self-energy as required.  However, this will not be true for $t \ne 0$.

The second issue is the dependence of the 5-D auto-correlation functions on the evolution times $t_1$, $t_2\ldots t_N$.  Order-by-order in perturbation theory all amplitudes decrease exponentially in the separation between the largest and smallest times measured in lattice units or oscillate with odd harmonics of the unit lattice frequency suggesting that critical slowing down has been eliminated.  However, the presence of polynomials in these times raises the possibility that when summed to all orders these polynomials may generate additional exponential time dependence that may cancel the large exponents present in the perturbative theory, replacing them by a weaker time dependence and causing a reemergence of critical slowing down, a superficially unlikely possibility.

Related to this question for our $\phi^4$ example, is the choice of the mass used in the Fourier-acceleration factor $(-\partial^2+\widetilde{m}^2)$ in Eqs.~\eqref{eq:L-FA1}, \eqref{eq:HMC-FA1} and \eqref{eq:L-FA-noise}.  For simplicity, in the discussion above we have used the bare mass $m_0$ in these factors.  This may be a poor choice since the Fourier-acceleration factor $q^2 + \widetilde{m}^2$ should be small compared to the lattice scale for physical $q^2$, something that will not be true if the bare mass $m_0 \sim 1/a$ is used in place of $\widetilde{m}$.  Apparently using the renormalized mass for $\widetilde{m}$ and in the unperturbed action would be a wiser choice.  In a gauge theory where the cut-off scale enters only logarithmically, such an issue may be less important.

\section{Conclusion and acknowledgement}

We have examined the stochastic evolution equations which can be used to describe the auto-correlation present in an ensemble of Monte Carlo samples generated using the Langevin and GHMC algorithms.  In both cases it appears that if Fourier acceleration is introduced the resulting dependence on the Langevin or GHMC simulation time is significantly changed.  The kernel used to examine auto-correlations in a perturbative expansion acquires a strong time dependence, decreasing or oscillating with a correlation time or frequency at the cutoff scale.  At least in perturbation theory this evolution time scale is unaffected by the longer times scales of the physical theory being studied indicating that critical slowing down has been eliminated.

We thank our colleagues from the RBC and UKQCD collaborations and the LatticeQCD ECP Application Development project for helpful discussions and Martin L\"uscher for raising the issue of HMC non-renormalizability which motivated this study.



\begin{thebibliography}{10}

\bibitem{Duane:1987de}
S.~Duane, A.~D. Kennedy, B.~J. Pendleton and D.~Roweth, \emph{{Hybrid Monte
  Carlo}}, \href{https://doi.org/10.1016/0370-2693(87)91197-X}{\emph{Phys.
  Lett.} {\bfseries B195} (1987) 216}.

\bibitem{Girolami:2011xx}
M.~Girolami and B.~Calderhead, \emph{{Riemann manifold Langevin and Hamiltonian
  Monte Carlo methods}}, {\emph{Journal of the Royal Statistical Society:
  Series B} {\bfseries 73} (2011) 123}.

\bibitem{Cossu:2017eys}
G.~Cossu, P.~Boyle, N.~Christ, C.~Jung, A.~JÃ¼ttner and F.~Sanfilippo,
  \emph{{Testing algorithms for critical slowing down}},
  \href{https://doi.org/10.1051/epjconf/201817502008}{\emph{EPJ Web Conf.}
  {\bfseries 175} (2018) 02008}
  [\href{https://arxiv.org/abs/1710.07036}{{\ttfamily 1710.07036}}].

\bibitem{Jin:2018xxx}
X.-Y. Jin, \emph{{Ensemble Quasi-Newton HMC}},  in \emph{{Proceedings, 36th
  International Symposium on Lattice Field Theory (Lattice 2018)}},
  vol.~LATTICE2018, p.~027, 2018.

\bibitem{Zhao:2018xxx}
Y.~Zhao, \emph{{Testing a new gauge-fixed Fourier acceleration algorithm}},  in
  \emph{{Proceedings, 36th International Symposium on Lattice Field Theory
  (Lattice 2018)}}, vol.~LATTICE2018, p.~156, 2018.

\bibitem{ZinnJustin:1986eq}
J.~Zinn-Justin, \emph{{Renormalization and Stochastic Quantization}},
  \href{https://doi.org/10.1016/0550-3213(86)90592-4}{\emph{Nucl. Phys.}
  {\bfseries B275} (1986) 135}.

\bibitem{Luscher:2011qa}
M.~Luscher and S.~Schaefer, \emph{{Non-renormalizability of the HMC
  algorithm}}, \href{https://doi.org/10.1007/JHEP04(2011)104}{\emph{JHEP}
  {\bfseries 04} (2011) 104} [\href{https://arxiv.org/abs/1103.1810}{{\ttfamily
  1103.1810}}].

\bibitem{Batrouni:1985jn}
G.~G. Batrouni, G.~R. Katz, A.~S. Kronfeld, G.~P. Lepage, B.~Svetitsky and
  K.~G. Wilson, \emph{{Langevin Simulations of Lattice Field Theories}},
  \href{https://doi.org/10.1103/PhysRevD.32.2736}{\emph{Phys. Rev.} {\bfseries
  D32} (1985) 2736}.

\bibitem{Parisi:1980ys}
G.~Parisi and Y.-s. Wu, \emph{{Perturbation Theory Without Gauge Fixing}},
  {\emph{Sci. Sin.} {\bfseries 24} (1981) 483}.

\bibitem{Horowitz:1991rr}
A.~M. Horowitz, \emph{{A Generalized guided Monte Carlo algorithm}},
  \href{https://doi.org/10.1016/0370-2693(91)90812-5}{\emph{Phys. Lett.}
  {\bfseries B268} (1991) 247}.

\end{thebibliography}

\providecommand{\href}[2]{#2}\begingroup\raggedright\endgroup

\end{document}